\documentclass[11pt]{article}
\title{Non-Existence of Black Hole Solutions for a Spherically 
Symmetric, Static Einstein-Dirac-Maxwell System}
\author{Felix Finster\thanks{Research supported by the Schweizerischer 
Nationalfonds. Mailing address after October 1998:
Max Planck Institute for Mathematics in the Sciences,
Inselstr.\ 22-26, 04103 Leipzig/Germany.},\ 
Joel Smoller\thanks{Research supported in part by the NSF, Grant No.\ 
DMS-G-9802370.}, and Shing-Tung Yau\thanks{Research supported in part 
by the NSF, Grant No.\ 33-585-7510-2-30.}}
\date{October 1998}

\newtheorem{Def}{Def.}[section]
\newtheorem{Thm}[Def]{Theorem}
\newtheorem{Prp}[Def]{Proposition}
\newtheorem{Lemma}[Def]{Lemma}

\newcommand{\Proof}{{\em{Proof: }}}
\newcommand{\QED}{\ \hfill $\FBox$ \\[1em]}

\newcommand{\spc}{\;\;\;\;\;\;\;\;\;\;}

\newcommand{\R}{\mbox{\rm I \hspace{-.8 em} R}}

\newcommand{\FBox}{\rule{2mm}{2.25mm}}

\begin{document}
\maketitle

\begin{abstract}
We consider for $j=\frac{1}{2}, \frac{3}{2},\ldots$ a spherically symmetric,
static system of $(2j+1)$ Dirac particles, each having total angular
momentum $j$. The Dirac particles interact via a classical 
gravitational and electromagnetic field.

The Einstein-Dirac-Maxwell equations for this system are derived.
It is shown that, under weak regularity conditions on the form of the
horizon, the only black hole solutions of the EDM equations are the
Reissner-Nordstr\"om solutions. In other words, the spinors must vanish
identically.
Applied to the gravitational collapse of a ``cloud'' of 
spin-$\frac{1}{2}$-particles to a black hole, our result indicates 
that the Dirac particles must eventually disappear inside the event 
horizon.
\end{abstract}

\section{Introduction}
In \cite{FSY1,FSY2}, particle-like solutions of the
Einstein-Dirac-Maxwell (EDM) equations were constructed for a static, spherically
symmetric singlet system. It was found that the 
solutions in a given state (i.e.\ in the ground state or in any fixed 
excited state) cease to exist if the rest mass $m$ of the fermions becomes 
larger than a certain threshold value $m_s$. The most natural physical 
interpretation of this observation is that for $m>m_s$, the 
gravitational interaction becomes so strong that a black hole would form. 
This suggests that there should be black hole solutions of the coupled EDM
equations for large fermion masses.
The work \cite{FSY3}, however, indicates that this intuitive picture 
of black hole formation is wrong. Namely, it was proved in 
\cite{FSY3} that the Dirac equation has no time-periodic 
solutions in a Reissner-Nordstr\"om black hole background, even if 
the Dirac particles have angular momentum and can thus, in the classical
picture, ``rotate around'' the black hole. This implies that if we neglect the
influence of the Dirac particles on the gravitational and electromagnetic
field, there are no black hole solutions of the EDM equations.

In order to understand if and how Dirac 
particles can form black holes, we study in this paper the 
fully coupled EDM equations. We do not assume the Dirac particles to be in
a spherically symmetric state; they are allowed to have angular momentum $j$. 
However, we arrange $2j+1$ of these particles in such a way that the total 
system is static and spherically symmetric. In the language of 
atomic physics, we consider the completely filled shell of states 
with angular momentum $j$. Classically, one can think of this 
multiple-particle system as of several Dirac particles rotating around a 
common center such that their angular momentum adds up to zero. Since 
the system of fermions is spherically symmetric, we get a 
consistent set of equations if we also assume spherical symmetry for 
the gravitational and electromagnetic field. This allows us to separate out
the angular dependence and reduce the problem to the analysis of a system 
of nonlinear ODEs.

We prove analytically that, under weak regularity conditions on the 
form of the horizon, the black hole solutions of our coupled EDM 
equations are either the Reissner-Nordstr\"om solutions (in which 
case the Dirac wave functions are identically zero), or the event 
horizon has the form of the extreme Reissner-Nordstr\"om metric. In 
the latter case, we show numerically that the Dirac wave functions 
cannot be normalized. Thus our Einstein-Dirac-Maxwell system does not 
admit black hole solutions.
Our results show that the study of black holes in the 
presence of Dirac spinors leads to unexpected physical effects.
Applied to the gravitational collapse of a ``cloud'' of Dirac particles,
this is a further indication that if an event horizon forms, the Dirac
particles must eventually disappear inside this horizon.

The methods used in this paper are quite different from those 
in \cite{FSY3}. Namely, in contrast to \cite{FSY3}, we do not derive
``matching conditions'' for the spinors across the horizon. We work 
here only with the equations outside the event horizon, and the proof 
relies on the nonlinear coupling of the spinors into the Einstein-Maxwell
equations.

\section{The Spherically Symmetric Multi-Particle System}
\setcounter{equation}{0}
The gravitational field is described by the static, spherically 
symmetric Lorentzian metric in polar coordinates
\begin{equation}
ds^2 \;=\; g_{ij} \:dx^i\:dx^j \;=\;
\frac{1}{T^2}\:dt^2 \:-\: \frac{1}{A} \:dr^2 \:-\: r^2 
\:d\vartheta^2 \:-\: r^2 \:\sin^2 \vartheta \:d\varphi^2
        \label{eq:1}
\end{equation}
with positive functions $A=A(r)$ and $T=T(r)$.
We consider as our space-time the region $r>\rho>0$ outside a ball of 
radius $\rho$ around the origin.
The physical situation we have in mind is that the surface 
$r=\rho$ is the event horizon of a black hole.
We assume the metric to be asymptotically Minkowskian,
\begin{equation}
        \lim_{r \rightarrow \infty} A(r) \;=\; 1 \;=\; \lim_{r \rightarrow 
        \infty} T(r) \;\;\; .
        \label{eq:21a}
\end{equation}
The electromagnetic field is described by a potential ${\cal{A}}$ of the 
form ${\cal{A}}=(-\phi, \vec{0})$, where $\phi$ is the Coulomb potential 
$\phi=\phi(r)$.

In direct generalization of the situation in the Reissner-Nordstr\"om 
background \cite{FSY3}, the Dirac operator $G$ takes the form
\begin{eqnarray}
G &=& i G^j(x) \:\frac{\partial}{\partial x^j} \:+\: B(x) \nonumber \\
&&\hspace*{-1.3cm}
=\; i T \gamma^0 \left(\frac{\partial}{\partial t} - i e \phi \right)
        + \gamma^r \left( i \sqrt{A} \frac{\partial}{\partial r} + \frac{i}{r} 
        \:(\sqrt{A}-1) -\frac{i}{2}\:\sqrt{A} \:\frac{T^\prime}{T} \right)
        + i \gamma^\vartheta \frac{\partial}{\partial \vartheta} + i
        \gamma^\varphi \frac{\partial}{\partial \varphi} \; , \;\;\;\;\;\;\;
        \label{eq:10}
\end{eqnarray}
where $\gamma^t$, $\gamma^r$, $\gamma^\vartheta$, and 
$\gamma^\varphi$ denote the $\gamma$-matrices of Minkowski space in 
polar coordinates,
\begin{eqnarray*}
\gamma^t &=& \gamma^0 \\
\gamma^r &=& \gamma^1 \:\cos \vartheta \:+\: \gamma^2 \:
        \sin \vartheta \: \cos \varphi \:+\: \gamma^3 \: \sin \vartheta \:
        \sin \varphi \\
\gamma^\vartheta &=& \frac{1}{r} \left( -\gamma^1 \:\sin \vartheta \:+\: \gamma^2 \:
        \cos \vartheta \: \cos \varphi \:+\: \gamma^3 \: \cos \vartheta \:
        \sin \varphi \right) \label{2.13aa} \\
\gamma^\varphi &=& \frac{1}{r \sin \vartheta} \left( -\gamma^2 \: \sin \varphi \:+\:
        \gamma^3 \: \cos \varphi \right) \;\;\; .
\end{eqnarray*}
As with the central force problem in Minkowski space \cite{S}, 
this Dirac operator commutes with: a) the time translation operator $i 
\partial_t$, b) the total angular momentum operator $J^2 = 
(\vec{L}+\vec{S})^2$, c) the $z$-component of total angular momentum 
$J_z$, and d) with the operator $\gamma^0 \:P$ (where $P$ is parity).
Since these operators also commute with each other, we can
write any solution of the Dirac equation as a linear 
combination of solutions which are simultaneous eigenstates of these operators.
We denote this ``eigenvector basis'' for the solutions by
\begin{equation}
\Psi^c_{jk \omega} \spc {\mbox{with}} \spc c=\pm, \;\;\;j=\frac{1}{2}, 
\frac{3}{2},\ldots, \;\;\; k=-j,-j+1,\ldots,j,\;\;\; \omega \in \R ;
\end{equation}
the eigenvalues are
\begin{eqnarray*}
i \partial_t \:\Psi^c_{jk \omega} & = & \omega \:\Psi^c_{jk \omega} \\
J^2 \: \Psi^c_{jk \omega} & = & j(j+1) \:\Psi^c_{jk \omega} \\
J_z \:\Psi^\pm_{jk \omega} &=& k \:\Psi^\pm_{jk \omega} \\
\gamma^0 P \:\Psi^\pm_{jk \omega} &=& \pm \Psi^\pm_{jk \omega} \:\times\:
\left\{ \begin{array}{cc} 1 & {\mbox{for $j+\frac{1}{2}$ even}} \\[.3em]
-1 & {\mbox{for $j+\frac{1}{2}$ odd}} \end{array} \right. \;\;\; .
\end{eqnarray*}
For the functions $\Psi^c_{jk \omega}$, the Dirac equation reduces to 
ordinary differential equations in the radial variable $r$.
The quantum number $k$ describes the orientation of the 
wave function $\Psi^c_{jk \omega}$ in space, and thus spherical symmetry 
of the Dirac operator implies that the radial Dirac equation is the same
for all values of $k$. In order to build up our many-particle system,
we take, for given $c$, $j$, and $\omega$, one solution of the radial 
Dirac equation and consider the system of the $2j+1$ particles
\[ \Psi^c_{jk \omega} \;\;\;,\spc k=-j,\ldots,j \]
corresponding to this radial solution. Using the formalism of 
many-particle quantum mechanics, we can describe the fermions with 
the Hartree-Fock state
\begin{equation}
        \Psi^{\mbox{\scriptsize{HF}}} \;=\; \Psi^c_{j \:k=-j\: \omega} \wedge
        \Psi^c_{j \:k=-j+1\: \omega} \wedge \cdots \wedge \Psi^c_{j \:k=j\: \omega}
        \;\;\; .
        \label{eq:HF}
\end{equation}
For simplicity, we usually avoid this formalism here; it is easier to 
just work with an orthonormal basis of the one-particle states.

For clarity, we point out that each fermion has non-zero angular 
momentum and is thus {\em{not}} in a spherically symmetric state. 
Nevertheless, the system of $2j+1$ particles is spherically 
symmetric; this can be verified in detail as follows: The Hartree-Fock 
state (\ref{eq:HF}) is an eigenstate of $J_z$; namely
\begin{eqnarray*}
J_z\: \Psi^{\mbox{\scriptsize{HF}}} &=& (J_z \:\Psi^c_{j \:k=-j\: 
\omega}) \wedge \Psi^c_{j \:k=-j+1\: \omega} \wedge \cdots \wedge
\Psi^c_{j \:k=j\: \omega} \\
&&+ \Psi^c_{j \:k=-j\: \omega} \wedge
        (J_z \:\Psi^c_{j \:k=-j+1\: \omega}) \wedge \cdots \wedge
        \Psi^c_{j \:k=j\: \omega} \\
&&+\: \cdots \:+\: \Psi^c_{j \:k=-j\: \omega} \wedge
        \Psi^c_{j \:k=-j+1\: \omega} \wedge \cdots \wedge
        (J_z \:\Psi^c_{j \:k=j\: \omega}) \\
&=& \sum_{k=-j}^j k \:\Psi^{\mbox{\scriptsize{HF}}} \;=\; 0 \;\;\; .
\end{eqnarray*}
Similarly, we can apply the ``ladder operators'' $J_\pm = J_x \pm i 
J_y$ to the Hartree-Fock state,
\begin{eqnarray}
J_\pm\: \Psi^{\mbox{\scriptsize{HF}}} &=& (J_\pm \:\Psi^c_{j \:k=-j\: 
\omega}) \wedge \Psi^c_{j \:k=-j+1\: \omega} \wedge \cdots \wedge
\Psi^c_{j \:k=j\: \omega} \nonumber \\
&&+ \Psi^c_{j \:k=-j\: \omega} \wedge
        (J_\pm \:\Psi^c_{j \:k=-j+1\: \omega}) \wedge \cdots \wedge
        \Psi^c_{j \:k=j\: \omega} \nonumber \\
&&+\: \cdots \:+\: \Psi^c_{j \:k=-j\: \omega} \wedge
        \Psi^c_{j \:k=-j+1\: \omega} \wedge \cdots \wedge
        (J_\pm \:\Psi^c_{j \:k=j\: \omega}) \;\;\; . \label{eq:J}
\end{eqnarray}
After substituting the relations
\[ J_\pm \: \Psi^c_{jk \omega} \;=\; \sqrt{j(j+1) - k(k \pm 1)} 
\:\Psi^c_{j \:k\pm1\: \omega} \;\;\; , \]
the anti-symmetry of the wedge product yields that each summand in 
(\ref{eq:J}) vanishes. We conclude that
\[ \vec{J} \: \Psi^{\mbox{\scriptsize{HF}}} \;=\; 0 \;\;\; . \]
Since the total angular momentum operator $\vec{J}$ is the 
infinitesimal generator of rotations, this implies that
$\Psi^{\mbox{\scriptsize{HF}}}$ is spherically symmetric.

For a physically meaningful solution of the Dirac equation, the wave 
function must be normalized. The normalization integral for the wave
functions $\Psi^c_{jk \omega}$ over the hypersurface
${\cal{H}}=\{t={\mbox{const}},\:r>\rho \}$ is
\begin{equation}
        \int_{\cal{H}} \overline{\Psi^c_{jk \omega}} G^j \Psi^c_{jk \omega} 
        \:\nu_j\:d\mu_{\cal{H}} \;\;\; ,
        \label{eq:13}
\end{equation}
where $\nu$ is the future-directed normal of ${\cal{H}}$, and where
$d\mu_{\cal{H}}$ denotes the invariant measure on ${\cal{H}}$ induced by the 
Lorentzian metric. For a normalized solution of the Dirac equation,
(\ref{eq:13}) gives the probability for the particle to be in the 
region $r>\rho$ outside the ball of radius $\rho$ centered at the origin.
It seems tempting to demand that, for a normalizable solution of the Dirac
equation, the integral (\ref{eq:13}) must be finite. However, as explained in 
\cite{FSY3}, the normalization integral inside the event horizon 
(which we do not consider here), is not necessarily positive, and it 
might happen that an infinite contribution near $r=\rho$ in 
(\ref{eq:13}) is compensated by an infinite negative contribution 
inside the horizon. Therefore we only demand that the normalization integral
away from the horizon is finite; namely
\begin{equation}
\int_{\{ t={\mbox{\scriptsize{const}}},\: r>r_0\}}
\overline{\Psi^c_{jk \omega}} G^j \Psi^c_{jk \omega} 
        \:\nu_j\:d\mu_{\cal{H}} \;<\; \infty \spc {\mbox{for every $r_0 > \rho$}}.
        \label{eq:37c}
\end{equation}

We remark that the singlet state of \cite{FSY1,FSY2} can be 
recovered from our multi-particle system by considering the case 
$j=\frac{1}{2}$.

\section{The EDM Equations}
\setcounter{equation}{0}
We now derive the system of differential equations. We begin by 
separating out the angular and time dependence in the Dirac equation. 
We choose the wave functions $\Psi^c_{jk \omega}$ of the previous 
section in the explicit form
\begin{eqnarray}
\Psi^+_{jk \omega} &=& e^{-i \omega t} \:\frac{\sqrt{T}}{r} \left( 
\begin{array}{c} \chi^k_{j-\frac{1}{2}} \:\Phi^+_{jk\omega \:1}(r) \\
i \chi^k_{j+\frac{1}{2}}\: \Phi^+_{jk\omega \:2}(r) \end{array} 
\right) \label{eq:3} \\
\Psi^-_{jk \omega} &=& e^{-i \omega t} \:\frac{\sqrt{T}}{r} \left( 
\begin{array}{c} \chi^k_{j+\frac{1}{2}} \:\Phi^-_{jk\omega \:1}(r) \\
i \chi^k_{j-\frac{1}{2}}\: \Phi^-_{jk\omega \:2}(r) \end{array} 
\right) \;\;\; , \label{eq:4}
\end{eqnarray}
where $\Phi^c_{jk \omega}$ are two-component radial functions, and 
where $\chi^k_{j \pm \frac{1}{2}}$, $j=\frac{1}{2}, 
\frac{3}{2},\ldots$, $k=-j,-j+1,\ldots, j$ denote the $2$-spinors
\begin{eqnarray*}
\chi^k_{j-\frac{1}{2}} &=& \sqrt{\frac{j+k}{2j}} 
\:Y^{k-\frac{1}{2}}_{j-\frac{1}{2}} \left( \begin{array}{c} 1 \\ 0 
\end{array} \right) \:+\: \sqrt{\frac{j-k}{2j}} 
\:Y^{k+\frac{1}{2}}_{j-\frac{1}{2}} \left( \begin{array}{c} 0 \\ 1
\end{array} \right) \\
\chi^k_{j+\frac{1}{2}} &=& \sqrt{\frac{j+1-k}{2j+2}} 
\:Y^{k-\frac{1}{2}}_{j+\frac{1}{2}} \left( \begin{array}{c} 1 \\ 0 
\end{array} \right) \:-\: \sqrt{\frac{j+1+k}{2j+2}} 
\:Y^{k+\frac{1}{2}}_{j+\frac{1}{2}} \left( \begin{array}{c} 0 \\ 1
\end{array} \right)
\end{eqnarray*}
($Y^m_l$ are the spherical harmonics). The functions 
$\chi^k_{j\pm\frac{1}{2}}$ are eigenvectors of the operator 
$K=\vec{\sigma} \vec{L} + 1$, 
\[ K \:\chi^k_{j\pm\frac{1}{2}} \;=\; \mp (j+\frac{1}{2})\: 
\chi^k_{j\pm\frac{1}{2}} \;\;\; . \]
Using the relations between the angular momentum operators (see 
\cite{FSY3} for details), the Dirac equation $(G-m) \Psi^c_{jk 
\omega}$ with $G$ as in (\ref{eq:10}) reduces to the ordinary 
differential equation
\begin{eqnarray}
\lefteqn{ \sqrt{A}\: \frac{d}{dr} \Phi^\pm_{jk \omega} } \nonumber \\
&=& \left[ (\omega - e \phi) T \left( 
\begin{array}{cc} 0 & -1 \\ 1 & 0 \end{array} \right) \:\pm\:
\frac{2j+1}{2r} \:\left( \begin{array}{cc} 1 & 0 \\ 0 & -1 \end{array} \right)
\:-\: m \:\left( \begin{array}{cc} 0 & 1 \\ 1 & 0 \end{array} 
\right) \right] \Phi^\pm_{jk \omega} \;\;\; .
        \label{eq:6}
\end{eqnarray}
Since this equation is independent of $k$, we will in the following 
omit the index $k$ and simply write $\Phi^c_{j \omega}$.
The normalization integral (\ref{eq:37c}) for the wave functions takes the form
\begin{equation}
        \int_{r_0}^\infty |\Phi^c_{j \omega}(r)|^2 \:\frac{\sqrt{T}}{A} 
        \:dr \;<\; \infty \spc {\mbox{for every $r_0 > \rho$}}.
        \label{eq:37}
\end{equation}

The Dirac equation (\ref{eq:6}) implies that the ``radial flux'' 
function (see \cite{FSY3})
\begin{equation}
        F(r) \;:=\; \overline{\Phi^c_{j \omega}(r)} \left( 
        \begin{array}{cc} 0 & -i \\ i & 0 \end{array} \right) \Phi^c_{j \omega}(r)
        \label{eq:15}
\end{equation}
is a constant, as is verified by a short computation. Since 
$|\Phi^c_{j \omega}|^2 \geq F$ and since the metric is asymptotically 
flat, the normalization integral (\ref{eq:13}) can be finite only if 
this constant is zero. We can thus assume that $F$ vanishes identically.
This means that the product $\overline{\Phi^c_{j\omega\: 1}} 
\Phi^c_{j \omega \:2}$ (constructed from the  two components of
$\Phi^c_{j \omega}$) is a real function. For a given radius 
$r_0 \in (\rho, \infty)$, we can thus arrange with a constant phase 
transformation
\[ \Phi^c_{j \omega}(r) \:\rightarrow\: e^{i \alpha} \:\Phi^c_{j 
\omega}(r) \;\;\;,\spc \alpha \in \R, \]
that both $\Phi^c_{j\omega \:1}(r_0)$ and $\Phi^c_{j\omega \:2}(r_0)$ are real.
Since all the coefficients in the Dirac equation (\ref{eq:6}) are 
real, it follows that the spinors $\Phi^c_{j\omega}(r)$ are real even for all
$r \in (\rho, \infty)$. We denote these two real components of
$\Phi^c_{j \omega}$ by $\alpha$ and $\beta$.

Next we must calculate the total current and energy-momentum tensor of 
the Dirac particles. With our explicit formulas (\ref{eq:3}) and 
(\ref{eq:4}) for the angular dependence of the wave functions, the 
anti-symmetrization in the Hartree-Fock state (\ref{eq:HF}) is 
trivial. One obtains that the electromagnetic current of the 
multi-particle system is simply the sum of the currents of all 
states $\Psi^c_{jk \omega}$,
\[ j^k \;=\; \overline{\Psi^{\mbox{\scriptsize{HF}}}} \:G^k\:
\Psi^{\mbox{\scriptsize{HF}}} \;=\; \sum_{k=-j}^j \overline{\Psi^c_{jk 
\omega}} \:G^k\: \Psi^c_{jk \omega} \;\;\; . \]
Because of spherical symmetry, the angular 
components $j^\vartheta$ and $j^\varphi$ of the Dirac current vanish. 
The ``sum rule''
\begin{equation}
\sum_{k=-j}^j \overline{\chi^k_{j \pm \frac{1}{2}}(\vartheta, \varphi)}
\: \chi^k_{j \pm \frac{1}{2}}(\vartheta, \varphi) \;=\; \frac{2j+1}{4 \pi}
\label{eq:12}
\end{equation}
yields, after a straightforward computation, that
\[ j^t(x) \;=\; \sum_{k=-j}^j \overline{\Psi^c_{j k \omega}} G^t(x)
\Psi^c_{j k \omega} \;=\; \frac{T^2}{r^2} \:(\alpha^2+\beta^2) 
\:\frac{2j+1}{4 \pi} \;\;\; . \]
To calculate the radial current $j^r$, we will need the formula 
\cite[Eqn.\ (3.7)]{FSY3}
\begin{equation}
\sigma^r \:\chi^k_{j \pm \frac{1}{2}} \;=\; \chi^k_{j \mp \frac{1}{2}} \;\;\; ,
        \label{eq:14}
\end{equation}
where $\sigma^r$ is the Pauli matrix in the radial direction,
\[ \sigma^r \;=\; \sigma^1 \:\cos \vartheta \:+\: \sigma^2 \:
        \sin \vartheta \: \cos \varphi \:+\: \sigma^3 \: \sin \vartheta \:
        \sin \varphi \;\;\; . \]
Using (\ref{eq:14}), (\ref{eq:12}), and the fact that the radial flux
$F(r)$ vanishes, we obtain
\[ j^r \;=\; \sum_{k=-j}^j \overline{\Psi^c_{j k \omega}} G^r(x)
\Psi^c_{j k \omega} \;=\; 0 \;\;\; . \]
Similar to the total current, the energy-momentum tensor of the
multi-particle system is simply the sum of the energy-momentum
of all states $\Psi^c_{jk \omega}$; it has the general form
\begin{equation}
        T_{ab} \;=\; \frac{1}{2} \:{\mbox{Re}} \sum_{k=-j}^j 
        \overline{\Psi^c_{jk \omega}} \left( i G_a \:(\partial_b - ie 
        {\cal{A}}_b) \:+\: i G_b \:(\partial_a - ie {\cal{A}}_a) \right)
        \Psi^c_{jk \omega}
        \label{eq:17}
\end{equation}
(this formula is obtained by varying the classical Dirac action; see 
e.g.\ \cite{FSY1}). The following calculation depends on whether the 
index $c$ of the wave functions in (\ref{eq:17}) is $c=+$ or $c=-$. We use the 
$\pm/\mp$-notation, whereby the upper and lower choice correspond to 
the cases $c=+$ and $c=-$, respectively.
From spherical symmetry, $T^t_\vartheta$, $T^t_\varphi$, $T^r_\vartheta$,
$T^r_\varphi$, and $T^\vartheta_\varphi$ must vanish. An easy computation
using the sum rule (\ref{eq:12}) and the Dirac equation (\ref{eq:6}) gives
\begin{eqnarray*}
T^t_t & = & \frac{(\omega - e \phi) \:T^2}{r^2} \:(\alpha^2 + 
\beta^2) \:\frac{2j+1}{4 \pi}  \\
T^r_r & = & -\frac{(\omega - e \phi) \:T^2}{r^2} \:(\alpha^2 + 
\beta^2) \:\frac{2j+1}{4 \pi} \:\pm\: \frac{T}{r^3} \:\alpha \beta\: 
\frac{(2j+1)^2}{4 \pi} \:+\: \frac{mT}{r^2} \:(\alpha^2-\beta^2) 
\:\frac{2j+1}{4 \pi} \;\;\; .
\end{eqnarray*}
The calculations of $T^\vartheta_\vartheta$ and $T^\varphi_\varphi$ 
are slightly more difficult. First, spherical symmetry yields that
\[ T^\vartheta_\vartheta \;=\; T^\varphi_\varphi \;=\; \frac{1}{2} 
\:{\mbox{Re}} \sum_{k=-j}^j \overline{\Psi^c_{jk \omega}} \:(i 
G^\vartheta \partial_\vartheta \:+\: i G^\varphi 
\partial_\varphi)\:\Psi^c_{jk \omega} \;\;\; . \]
The formula
\[ G^\vartheta \partial_\vartheta + G^\varphi \partial_\varphi \;=\; 
-\frac{1}{r} \:\sigma^r\:\vec{\gamma} \vec{L} \]
allows us to rewrite the angular derivatives using the angular 
momentum operator $\vec{L}$. This gives
\begin{eqnarray*}
T^\vartheta_\vartheta \;=\; T^\varphi_\varphi &=&
\frac{\alpha \beta}{2} \:\frac{T}{r^3} \:{\mbox{Re}} \sum_{k=-j}^j 
\left( \overline{\chi^k_{j \mp \frac{1}{2}}} (\sigma^r 
\:\vec{\sigma} \vec{L}) \chi^k_{j \pm \frac{1}{2}} \:-\:
\overline{\chi^k_{j \pm \frac{1}{2}}} (\sigma^r 
\:\vec{\sigma} \vec{L}) \chi^k_{j \mp \frac{1}{2}} \right) \\
&\stackrel{(\ref{eq:14})}{=}&
\frac{\alpha \beta}{2} \:\frac{T}{r^3} \:{\mbox{Re}} \sum_{k=-j}^j 
\left( \overline{\chi^k_{j \pm \frac{1}{2}}} \:\vec{\sigma} \vec{L}\:
\chi^k_{j \pm \frac{1}{2}} \:-\:
\overline{\chi^k_{j \mp \frac{1}{2}}} \:\vec{\sigma} \vec{L}\:
\chi^k_{j \mp \frac{1}{2}} \right) \;\;\; .
\end{eqnarray*}
We now use the fact that the spinors $\chi^k_{j \pm \frac{1}{2}}$ are 
eigenvectors of the operator $\vec{\sigma} \vec{L}=K-1$, and carry out 
the $k$-summation with (\ref{eq:12}) to get
\begin{eqnarray*}
T^\vartheta_\vartheta \;=\; T^\varphi_\varphi &=&
\mp \frac{\alpha \beta}{2} \:\frac{T}{r^3} \left( 
(j+\frac{3}{2})+(j-\frac{1}{2}) \right) \frac{2j+1}{4 \pi}
\;=\; \mp \frac{T}{r^3} \:\alpha \beta\: \frac{(2j+1)^2}{8 \pi} \;\;\; .
\end{eqnarray*}

Finally, we put the obtained formulas for the Dirac current and 
energy-momentum tensor together with the Maxwell energy-momentum
tensor \cite{FSY2} into the Einstein and Maxwell equations
\[ R^i_j - \frac{1}{2} \:R\: \delta^i_j \;=\; -8 \pi \:T^i_j 
\;\;\;,\spc \nabla_l F^{kl} \;=\; 4 \pi e \:j^k \;\;\; . \]
The Einstein equations reduce to two first-order equations for $A$ 
and $T$; similarly, Maxwell's equations simplify to one second-order 
equation, and we end up with the following system of EDM equations:
\begin{eqnarray}
\sqrt{A} \:\alpha^\prime &=& \pm \frac{2j+1}{2 r} \:\alpha \:-\: 
        ((\omega-e \phi) T + m) \:\beta \label{eq:E1} \\
\sqrt{A} \:\beta^\prime  &=& ((\omega-e \phi) T - m) \:\alpha \:\mp\:
        \frac{2j+1}{2 r} \:\beta \label{eq:E2} \\
r \:A^\prime & = & 1-A \:-\: 2(2j+1) (\omega-e \phi) T^2
        \:(\alpha^2 + \beta^2) \:-\: r^2 A T^2\: |\phi^\prime|^2
        \label{eq:E3} \\
2 r A \:\frac{T^\prime}{T} &=& A-1 \:-\: 2(2j+1) (\omega-e \phi) T^2 
        \:(\alpha^2+\beta^2) \:\pm\: 2 \:\frac{(2j+1)^2}{r} \:T\: \alpha \beta 
        \nonumber \\
&&+2(2j+1) \:m T \:(\alpha^2-\beta^2) \:+\: r^2 A T^2\: |\phi^\prime|^2
        \label{eq:E4} \\
r^2 A \:\phi^{\prime \prime} &=& -(2j+1)\: e \:(\alpha^2+\beta^2)
        -  \left( 2r A  + r^2 A \:\frac{T^\prime}{T}
        + \frac{r^2}{2} \:A^\prime \right) \phi^\prime \;\;\;\; .
        \label{eq:E5}
\end{eqnarray}
The two cases for the signs $\pm/\mp$ correspond to the two values
$c=\pm$ for the fermionic wave functions $\Psi^c_{jk \omega}$.
Equations (\ref{eq:E1}) and (\ref{eq:E2}) are the Dirac equations 
(\ref{eq:6}). The Einstein equations are (\ref{eq:E3}) and 
(\ref{eq:E4}), and (\ref{eq:E5}) is Maxwell's equation. According to
(\ref{eq:37}), the normalization condition is
\begin{equation}
        \int_{r_0}^\infty (\alpha^2 + \beta^2) \:\frac{\sqrt{T}}{A} 
        \:dr \;<\; \infty \spc {\mbox{for every}}\spc r_0 > \rho.
        \label{eq:nc}
\end{equation}

We remark that the system (\ref{eq:E1})--(\ref{eq:E5}), (\ref{eq:nc}) 
has particle-like solutions, which can be constructed numerically 
using the methods in \cite{FSY1}. The mass-energy spectrum of the 
solutions has the same qualitative behavior as for the two-particle 
EDM system \cite{FSY2}.

\section{Non-Existence Results}
\setcounter{equation}{0}
We want to investigate {\em{black hole solutions}} of the system
(\ref{eq:E1})--(\ref{eq:E5}). This means, more precisely, that we consider
solutions of (\ref{eq:E1})--(\ref{eq:E5}) defined outside the ball of
radius $\rho$ around the origin which are asymptotically 
flat, (\ref{eq:21a}), and satisfy the normalization condition 
(\ref{eq:37}). We assume that the surface $r=\rho$ is an event horizon; i.e.\
the function $A(r)$ tends to zero for $r \searrow \rho$,
whereas $T(r)$ goes to 
infinity in this limit. We make the following assumptions on the form 
of the horizon:
\begin{description}
\item[{\rm{(I)}}] The volume element $\sqrt{| {\mbox{det }} g_{ij}|} = r^2 
A^{-\frac{1}{2}}\: T^{-1}$ is smooth and non-zero on the horizon, i.e.
\[ T^{-2} \:A^{-1},\: T^2 A \;\in\; C^\infty([\rho, \infty)) \;\;\; . \]
This assumption is sometimes called: {\em{the horizon is regular}}.
\item[{\rm{(II)}}] The strength of the electromagnetic field is 
given by the scalar $F_{ij} \:F^{ij} = -2 |\phi^\prime|^2 \:A\:T^2$
with the electromagnetic field tensor $F_{ij} = 
\partial_i {\cal{A}}_j - \partial_j {\cal{A}}_i$.
We assume this scalar to be bounded near the horizon; thus in view of (I) we
assume that
\[ |\phi^\prime(r)| \;<\; c_1 \;\;\;,\spc \rho<r<\rho+\varepsilon_1 \]
for some positive constants $c_1$, $\varepsilon_1$.
\item[{\rm{(III)}}] The function $A(r)$ obeys a power law, i.e.
\begin{equation}
        A(r) \;=\; c \:(r-\rho)^s \:+\: {\cal{O}}((r-\rho)^{s+1}) 
        \;\;\;,\spc r>\rho
        \label{eq:B}
\end{equation}
with positive constants $c$ and $s$.
\end{description}
If assumptions (I) or (II) were violated, an
observer freely falling into the black hole
would feel strong forces when crossing the horizon.
Assumption (III) is a technical condition which seems general enough to
include all physically relevant horizons. For example, the 
Schwarzschild horizon has $s=1$, whereas $s=2$ corresponds to the 
extreme Reissner-Nordstr\"om horizon. However, assumption (III) does not seem
to be essential for the statement of our non-existence results; with more
mathematical effort, it could be weakened or perhaps even omitted completely.
We now state our main result:
\begin{Thm}
\label{thm1}
The black hole solutions of the EDM system 
(\ref{eq:E1})--(\ref{eq:E5}) satisfying the regularity conditions 
(I), (II), and (III) either coincide with a non-extreme Reissner-Nordstr\"om 
solution with $\alpha=0=\beta$, or $s=2$ and the following asymptotic
expansions hold near $r=\rho$:
\begin{eqnarray}
A(r) &=& A_0 \:(r-\rho)^2 \:+\: {\cal{O}}((r-\rho)^3) \label{eq:p1} \\
T(r) &=& T_0 \:(r-\rho)^{-1} \:+\: {\cal{O}}((r-\rho)^0) 
\label{eq:p2} \\
\phi(r) &=& \frac{\omega}{e} \:+\: \phi_0 \:(r-\rho)
\:+\: {\cal{O}}((r-\rho)^2) \label{eq:p3} \\
\alpha(r) &=& \alpha_0 \:(r-\rho)^\kappa \:+\:
{\cal{O}}((r-\rho)^{\kappa+1}) \label{eq:p4a} \\
\beta(r) &=& \beta_0 \:(r-\rho)^\kappa \:+\:
{\cal{O}}((r-\rho)^{\kappa+1}) \label{eq:p4}
\end{eqnarray}
with positive constants $A_0$, $T_0$ and real parameters 
$\phi_0$, $\alpha_0$, $\beta_0$.
The power $\kappa$ must satisfy the constraint
\begin{equation}
        \frac{1}{2} \;<\; \kappa \;=\; \frac{1}{A_0} \:\sqrt{m^2 - e^2 
        \phi_0^2 T_0^2 + \left( \frac{2j+1}{2 \rho} \right)^2} \;\;\; ,
        \label{eq:p5}
\end{equation}
and the spinor coefficients $\alpha_0$ and $\beta_0$ are related by
\begin{equation}
        \alpha_0 \left( \sqrt{A_0} \:\kappa \:\pm\: \frac{2j+1}{2 \rho} 
        \right) \;=\; -\beta_0 \:(m - e \phi_0 T_0) \;\;\; ,
        \label{eq:p6}
\end{equation}
where `$\pm$' refers to the two choices of the signs in
(\ref{eq:E1})--(\ref{eq:E5}).
\end{Thm}
We begin the analysis with the case that the power $s$ in (\ref{eq:B}) 
is less than 2.
\begin{Lemma}
\label{lemma1}
Assume that $s<2$ and that $(\alpha, \beta, A, T, \phi)$ is a black-hole 
solution where the spinors $(\alpha, \beta)$ are not identically zero.
Then the function $(\alpha^2 + \beta^2)$ is bounded both from above 
and from below near the horizon, i.e.\ there are constants $c, 
\varepsilon>0$ with
\begin{equation}
        c \;\leq\; \alpha(r)^2 + \beta(r)^2 \;\leq\; \frac{1}{c} 
        \;\;\;,\spc \rho < r < \rho+\varepsilon \;\;\; .
        \label{eq:n7}
\end{equation}
\end{Lemma}
{\Proof}
The Dirac equations (\ref{eq:E1}),(\ref{eq:E2}) imply that
\begin{eqnarray}
\sqrt{A} \:\frac{d}{dr} (\alpha^2 + \beta^2) &=& 2
\left( \begin{array}{c} \alpha \\ \beta \end{array} \right)^T
\left( \begin{array}{cc} \displaystyle \pm\frac{2j+1}{2r} & -m \\
-m & \displaystyle \mp\frac{2j+1}{2r} \end{array} \right)
\left( \begin{array}{c} \alpha \\ \beta \end{array} \right) \nonumber \\
&\leq& \left(4 m^2 + \frac{(2j+1)^2}{r^2} \right)^{\frac{1}{2}} (\alpha^2 + 
\beta^2) \;\;\; . \label{eq:n3}
\end{eqnarray}
Since $(\alpha, \beta)$ is a non-trivial solution, the uniqueness 
theorem for solutions of ODEs implies that $(\alpha^2 + \beta^2)(r)$ 
is non-zero for all $\rho < r < \rho+\varepsilon$. Thus we can divide 
Eq.\ (\ref{eq:n3}) by $\sqrt{A} \:(\alpha^2 + \beta^2)$ and 
integrate. This yields the bound
\begin{eqnarray}
\lefteqn{ \left| \log ((\alpha^2 + \beta^2)(\rho + \varepsilon)) \:-\:
\log ((\alpha^2 + \beta^2)(r)) \right| } \nonumber \\
&\leq& \int_r^{\rho + \varepsilon}
A^{-\frac{1}{2}}(t) \left(4 m^2 + \frac{(2j+1)^2}{t^2} 
\right)^{\frac{1}{2}} dt \;\;\;  .
        \label{eq:n4}
\end{eqnarray}
Since $s<2$, we see that the integrand in (\ref{eq:n4}) is 
integrable at $r=\rho$. Thus the right-hand side of (\ref{eq:n4}) is
majorized by
\[ \int_\rho^{\rho + \varepsilon}
A^{-\frac{1}{2}}(t) \left(4 m^2 + \frac{(2j+1)^2}{t^2} 
\right)^{\frac{1}{2}} dt \;\;\; , \]
so we may take the limit $r \searrow \rho$
in (\ref{eq:n4}) to obtain the estimate (\ref{eq:n7}).
\QED
\begin{Prp}
\label{prp1}
For $0<s<2$, the only black hole solutions of the system 
(\ref{eq:E1})--(\ref{eq:E5}) are the non-extreme Reissner-Nordstr\"om 
solutions.
\end{Prp}
{\Proof}
We shall assume that we are given a black hole solution which is not the 
Reissner-Nordstr\"om solution, and obtain a contradiction. In this 
case, the spinors $(\alpha, \beta)$ are 
not identically zero, so we may apply Lemma \ref{lemma1} and conclude 
that the spinors are bounded near $r=\rho$.

We first consider the differential equation for $AT^2$. The Einstein 
equations (\ref{eq:E3}) and (\ref{eq:E4}) give
\begin{eqnarray}
r \:\frac{d}{dr}(A T^2) &=& -4(2j+1) \:(\omega - e \phi)
\:T^4\:(\alpha^2+\beta^2) \:\pm\: 2 \:\frac{(2j+1)^2}{r} 
\:T^3\:\alpha \beta \nonumber \\
&&+2 (2j+1) \:m \:T^3\: (\alpha^2 - \beta^2) \;\;\; .
\label{eq:nca}
\end{eqnarray}
According to the regularity condition (I), the left and thus also the 
right side of this equation is smooth. Since the spinors are bounded 
away from zero near $r=\rho$, and since $T \rightarrow \infty$ as $r 
\searrow \rho$, we see that
\begin{equation}
\lim_{\rho < r \rightarrow \rho} (\omega - e \phi(r)) \;=\; 0 \;\;\; .
        \label{eq:n8}
\end{equation}

We write Maxwell's equation (\ref{eq:E5}) in the form
\begin{equation}
\phi^{\prime \prime} \;=\; -\frac{1}{A} \:\frac{(2j+1)\: e}{r^2}
\:(\alpha^2+\beta^2) \:-\: \frac{1}{r^2 \:\sqrt{A} \:T} \:[r^2 
\:\sqrt{A} \:T]^\prime \:\phi^\prime \;\;\; .
        \label{eq:n5}
\end{equation}
According to the regularity condition (I), the square 
bracket in (\ref{eq:n5}), and thus the whole coefficient of 
$\phi^\prime$, is a smooth function. However, the factor $A^{-1}$ in 
the first summand in (\ref{eq:n5}) blows up on the horizon.
If $s \geq 1$, the singularity of $A^{-1}$ is not integrable. This 
implies that $|\phi^\prime|$ is unbounded on the horizon,
contradicting the regularity condition (II). We conclude that $s<1$.
We can then integrate Eq.\ (\ref{eq:n5}) and obtain the 
local expansion around the horizon
\[ \phi^\prime(r) \;=\; c_1 \:(r-\rho)^{-s+1} \:+\: c_2 \:+\: 
{\cal{O}}((r-\rho)^{-s+2}) \;\;\; , \]
where $c_2$ is an integration constant.
Integrating once again and using (\ref{eq:n8}) yields the expansion
\begin{equation}
\phi(r) \;=\; c_1 \:(r-\rho)^{-s+2} \:+\: c_2 \:(r-\rho) \:+\: 
\frac{\omega}{e} \:+\: {\cal{O}}((r-\rho)^{-s+3}) \;\;\; .
        \label{eq:n6}
\end{equation}

Finally, we substitute the expansion (\ref{eq:n6}) into the $A$-equation
(\ref{eq:E3}). Since the functions $A$ and $r^2 A T^2 
|\phi^\prime|^2$ are bounded near the horizon, and since (\ref{eq:n6}) 
implies that $(\omega - e \phi) = {\cal{O}}(r-\rho)$, whereas $T^2 
(\alpha^2 + \beta^2) \sim (r-\rho)^{-s}$ with $s<1$, the right side 
of (\ref{eq:E3}) is bounded in the limit $r \searrow \rho$. 
However, the left side diverges like $r A^\prime(r) \sim (r-\rho)^{s-1}$.
This is a contradiction.
\QED
It remains to consider the case $s \geq 2$.
\begin{Lemma}
\label{lemma3}
If $s \geq 2$,
\begin{equation}
\lim_{\rho < r \rightarrow \rho} (r-\rho)^{-\frac{s}{2}} 
\:(\alpha^2 + \beta^2)(r) \;=\; 0 \;\;\; .
        \label{eq:n9}
\end{equation}
\end{Lemma}
{\Proof}
As in the proof of Proposition \ref{prp1}, we consider the 
Maxwell equation (\ref{eq:n5}). Since $|\phi^\prime|$ is bounded near 
the horizon according to condition (II), we conclude from (I) that the 
inhomogeneous term in (\ref{eq:n5}) must be integrable,
\begin{equation}
\int_\rho^{\rho + \varepsilon} \frac{1}{A} \:(\alpha^2 + \beta^2) 
\;<\; \infty \;\;\; .
        \label{eq:na}
\end{equation}

Next we take the derivative of the function in (\ref{eq:n9}),
\[ \frac{d}{dr} \left( (r-\rho)^{-\frac{s}{2}} \:(\alpha^2 + 
\beta^2) \right) \;=\; -\frac{s}{2} \:(r-\rho)^{-\frac{s}{2}-1} 
\:(\alpha^2 + \beta^2) \:+\: (r-\rho)^{-\frac{s}{2}} \:\frac{d}{dr} 
(\alpha^2 + \beta^2) \;\;\; , \]
and substitute the bound (\ref{eq:n3}),
\begin{eqnarray}
\lefteqn{ \left| \frac{d}{dr} \left( (r-\rho)^{-\frac{s}{2}} \:(\alpha^2 + 
\beta^2) \right) \right| } \nonumber \\
&\leq& \frac{s}{2} \:(r-\rho)^{-\frac{s}{2}-1} \:(\alpha^2+\beta^2) \:+\:
\left(4m^2 + \frac{(2j+1)^2}{r^2} \right)^{\frac{1}{2}}\: A^{-\frac{1}{2}} 
\:(r-\rho)^{-\frac{s}{2}} \:(\alpha^2 + \beta^2) \;\; . \spc
\label{eq:428}
\end{eqnarray}
Since $s \geq 2$, we have $(r-\rho)^{-\frac{s}{2}-1}<(r-\rho)^{-s}$, 
and thus (\ref{eq:na}) implies that the first summand on the right side of
(\ref{eq:428}) is integrable. Using
\[ A^{-\frac{1}{2}} \:(r-\rho)^{-\frac{s}{2}} \:(\alpha^2+\beta^2) \;=\;
A^{-1} \:(\alpha^2+\beta^2) \:\left[ A^{\frac{1}{2}} \:
(r-\rho)^{-\frac{s}{2}} \right] \;=\; {\cal{O}}(A^{-1}
\:(\alpha^2+\beta^2)) \;\;\; , \]
we see that, according to (\ref{eq:na}), the second summand in
(\ref{eq:428}) is also integrable. As a consequence, the function
$(r-\rho)^{-\frac{s}{2}}\: (\alpha^2+\beta^2)$ has a limit at $r=\rho$. 
If this limit were non-zero, the integral (\ref{eq:na}) would diverge.
We conclude that this limit must be zero.
\QED

\begin{Lemma}
\label{lemma4}
If $s \geq 2$, the function $|\phi^\prime|$ has a finite, non-zero 
limit on the horizon; namely
\begin{equation}
        \lim_{\rho<r \rightarrow \rho} |\phi^\prime| \;=\; 
        \frac{1}{\rho} \:\lim_{\rho<r \rightarrow \rho} A^{-\frac{1}{2}} 
        \:T^{-1} \;>\; 0 \;\;\; .
        \label{eq:G}
\end{equation}
\end{Lemma}
{\Proof}
To begin, we first show that
\begin{equation}
\lim_{\rho<r \rightarrow \rho} (\omega - e \phi) \:T^2 \: (\alpha^2 + 
\beta^2) \;=\; 0 \;\;\; .
        \label{eq:nb}
\end{equation}
If the function $|(\omega - e \phi)T|$ is bounded, then (\ref{eq:nb}) 
is an immediate consequence of Lemma \ref{lemma3}. Thus we must only 
consider the case that $|(\omega - e \phi)T|$ is unbounded near the 
horizon. The differential equation for $AT^2$, (\ref{eq:nca}), gives 
the estimate
\[ \left| r \:\frac{d}{dr} A T^2 \right| \;\geq\; T^3 \:(\alpha^2 + 
\beta^2) \left( 4(2j+1) \:|(\omega-e \phi)T| \:-\: 2 
\:\frac{(2j+1)^2}{r} \:-\: 2(2j+1) \:m \right) \;\;\;. \]
According to assumption (I), the left side of this inequality is bounded 
near the horizon. Using that $|(\omega - e \phi)T|$ becomes 
arbitrarily large near the horizon, we conclude that the function 
$T^3 \:(\alpha^2+\beta^2) \:|(\omega - e \phi)T|$ must be bounded. This 
implies (\ref{eq:nb}).

We now consider the $A$-equation (\ref{eq:E3}). Since $s \geq 2$, the left 
side of (\ref{eq:E3}) converges to zero for $r \searrow \rho$.
Thus the right side of (\ref{eq:E3}) must also go to zero in this 
limit,
\[ 0 \;=\; \lim_{\rho<r \rightarrow \rho} 1-A-2(2j+1)(\omega - e 
\phi) \:T^2 \:(\alpha^2 + \beta^2) - r^2 \:A\:T^2\:|\phi^\prime(r)|^2 \;\;\; . \]
Using (\ref{eq:nb}) completes the proof.
\QED
We can now rule out the case $s>2$; namely we have
\begin{Prp}
\label{prp2}
For $s>2$, there are no solutions of the system (\ref{eq:E1})--(\ref{eq:E5}).
\end{Prp}
{\Proof}
According to Lemma \ref{lemma4}, the function $(\omega - e\phi)$ has 
a Taylor expansion around the horizon with non-vanishing linear term,
\[ (\omega - e \phi)(r) \;=\; c \:+\: d \:(r - \rho) \:+\: o(r-\rho) 
\spc {\mbox{with}} \spc |d| \;=\; \frac{e}{\rho} \:\lim_{\rho<r 
\rightarrow \rho} A^{-\frac{1}{2}} \:T^{-1} \;\;\; . \]
Thus the coefficient $(\omega - e \phi)T$ in the Dirac equation 
(\ref{eq:E1}), (\ref{eq:E2}) is monotone near the horizon and 
diverges. Using (\ref{eq:E1}) and (\ref{eq:E2}), this implies
that the vector $(\alpha, \beta)$ spins around the origin faster and
faster as $r$ approaches the 
horizon, which suggests that it cannot go to zero in this limit.
In fact, \cite[Lemma 5.1]{FSY3} yields that the spinors are bounded away
from zero,
\[ \liminf_{\rho<r \rightarrow \rho} \:(\alpha^2+\beta^2) \;>\; 0 
\;\;\; . \]
This contradicts Lemma \ref{lemma3}.
\QED\\[.5em]
{\em{Proof of Theorem \ref{thm1}: }}
According to Proposition \ref{prp1} and Proposition \ref{prp2}, we
must only consider the case $s=2$; thus (\ref{eq:p1}) and 
(\ref{eq:p2}) hold. Lemma \ref{lemma3} yields that the function $(\alpha^2 + 
\beta^2)$ goes to zero on the horizon. Applying \cite[Lemma 
5.1]{FSY3}, one sees that the function $(\omega - e \phi) T$ cannot 
diverge monotonically near the horizon. On the other hand, Lemma 
\ref{lemma4} shows that $(\omega - e \phi)$ has a Taylor expansion 
around the horizon with non-vanishing linear term. We conclude that
(\ref{eq:p3}) holds, and that $(\omega - e \phi)T$ has a finite
limit on the horizon,
\[ \lim_{\rho < r \rightarrow \rho} (\omega - e \phi)T \;=\; \lambda 
\spc {\mbox{with}} \spc |\lambda| \;\stackrel{(\ref{eq:G})}{=}\;
\frac{e}{\rho} \:\lim_{\rho<r 
\rightarrow \rho} (r-\rho)^{-1} \:A^{-\frac{1}{2}} \;\;\; . \]
Exactly as in \cite[Section 5]{FSY3}, one can rewrite the radial Dirac 
equations as ODEs in the variable
\[ u(r) \;=\; -r-\rho \:\ln(r-\rho) \]
and apply the stable manifold theorem \cite[Thm.\ 4.1]{C} to conclude
that $\alpha$ and $\beta$ satisfy a power law near the horizon,
(\ref{eq:p4a}) and (\ref{eq:p4}). Lemma \ref{lemma3} gives the 
bound $\kappa>\frac{1}{2}$.

Now we substitute the expansions (\ref{eq:p1})--(\ref{eq:p4}) into 
the Dirac equations (\ref{eq:E1}),(\ref{eq:E2}). This gives the 
conditions
\begin{eqnarray*}
\sqrt{A_0} \:\kappa\:\alpha_0 &=& \pm\frac{2j+1}{2 \rho} \:\alpha_0 
\:+\: (e \phi_0 T_0 - m) \:\beta_0 \\
\sqrt{A_0} \:\kappa\:\beta_0 &=& -(e \phi_0 T_0 + m) \alpha_0
\:\mp\: \frac{2j+1}{2 \rho} \:\beta_0 \;\;\; ,
\end{eqnarray*}
which are equivalent to (\ref{eq:p5}) and (\ref{eq:p6}).
\QED

Our main theorem gives strong restrictions for the behavior of black hole 
solutions near the event horizon. According to the condition 
$\kappa>\frac{1}{2}$, the Dirac wave functions must decay so fast
in the limit $r \searrow \rho$ that they have no influence on the asymptotic
form of both the metric and the electric field on the horizon. Namely,
according to (\ref{eq:p1})--(\ref{eq:p3}), the metric and electric field must
behave near the horizon like a vacuum solution, more precisely like the
extreme Reissner-Nordstr\"om solution. The restriction to the extremal
case means physically 
that the electric charge of the black hole must be so large that the 
electric repulsion balances the gravitational attraction and prevents 
the Dirac particles from ``falling into'' the black hole. This is 
certainly not the physical situation which one can expect in the 
gravitational collapse of e.g.\ a star in the Universe. Nevertheless, 
extreme Reissner-Nordstr\"om black holes have zero temperature \cite{H} 
and can thus be considered as the asymptotic states of black holes
emitting Hawking radiation. For this reason, it is interesting to study
if the asymptotic expansion of Theorem \ref{thm1} leads to global 
solutions of the EDM equations.

For an extreme Reissner-Nordstr\"om background field, it is proven in 
\cite[Section 5]{FSY3} that the solutions of the Dirac equation 
satisfying (\ref{eq:p4a}),(\ref{eq:p4}) violate the normalization condition 
(\ref{eq:nc}). Thus the question is if the influence of the spinors on 
the gravitational and electromagnetic fields can make the 
normalization integral finite. This is a hard analytic problem, because one 
must control the global behavior of the solution. However, we have done
numerical investigations, taking the expansions in Theorem 
\ref{thm1} as initial condition on the horizon and solving the 
equations for increasing $r$.
It turns out that the solutions either develop singularities for 
finite $r$, or the spinors $(\alpha, \beta)$ are not normalizable.
Thus our numerics show that the expansions in Theorem \ref{thm1} 
do not give normalizable black hole solutions of the EDM equations.

\begin{tabular}{ll}
\\
Mathematics Department, & Mathematics Department,\\
Harvard University, & The University of Michigan,\\
Cambridge, MA 02138  \hspace*{.5cm}(FF \& STY)\hspace*{1cm}
& Ann Arbor, MI 48109 \hspace*{.5cm} (JS)\\
\\
\end{tabular}

{\tt{felix@math.harvard.edu, smoller@umich.edu, yau@math.harvard.edu}}

\end{document}